\documentclass[journal]{IEEEtran}
\usepackage{cite}
\usepackage{amsmath,amssymb,amsfonts}
\usepackage{algorithmic}
\usepackage{xcolor}
\usepackage{pgfplots}
\usepackage{float}
\usepackage{tikz}
\usepackage{flushend}
\usepackage{caption}
\usepackage{booktabs}
\usepackage{colortbl}
\usepackage{tabularray}
\usepackage{multirow} 
\usepackage{nicematrix}
\usepackage{graphicx,subcaption}
\usepackage{lipsum,multicol}
\usepackage{hyperref}
\usepackage{orcidlink}
\usepackage[ruled,linesnumbered]{algorithm2e} % Required for typesetting algorithms

\definecolor{cite-blue}{HTML}{008ADA}
\definecolor{gainsboro229}{RGB}{229,229,229}
\definecolor{tablecolor}{RGB}{49, 131, 245}
\definecolor{tab1}{HTML}{f7fdff}
\definecolor{tab2}{HTML}{DEF0F9}
\definecolor{tab3}{HTML}{CFE9F7}
\definecolor{own_green}{HTML}{009900}

\definecolor{lightgray}{RGB}{240, 240, 240}
\definecolor{steelblue}{RGB}{52,138,189}
\definecolor{tab2}{HTML}{DEF0F9}
\tikzset{%
  dots/.style args={#1per #2}{%
    line cap=round,
    dash pattern=on 0 off #2/#1
  }
}

\hypersetup{
    colorlinks=true,
    citecolor=red,
    linkcolor=cite-blue,
    urlcolor=cite-blue
}
\usepackage{textcomp}
\def\BibTeX{{\rm B\kern-.05em{\sc i\kern-.025em b}\kern-.08em
    T\kern-.1667em\lower.7ex\hbox{E}\kern-.125emX}}
\begin{document}
\title{Optimizing Universal Lesion Segmentation: State Space Model-Guided Hierarchical Networks with Feature Importance Adjustment}

\author{Kazi Shahriar Sanjid\textsuperscript{\orcidlink{0009-0001-6845-8881}},
Md. Tanzim Hossain\textsuperscript{\orcidlink{0000-0002-4776-7220}}, Md. Shakib Shahariar Junayed\textsuperscript{\orcidlink{0009-0009-8755-6370}} %\IEEEmembership{Student Member, IEEE}, 
and
M. Monir Uddin\textsuperscript{\orcidlink{0000-0002-9817-6156}}
\thanks{This work was funded by the North South University Conference Travel and Research Grants (NSU-CTRG) under Grant No. CTRG-22-SEPS-06}
\thanks{The first three authors contributed equally to this work. \textit{Asterisk indicates corresponding author}}
\thanks{Kazi Shahriar Sanjid, Md. Tanzim Hossain and Md. Shakib Shahariar Junayed are with the Department of Electrical \& Computer Engineering, North South University, Dhaka-1229, Bangladesh (e-mail: kazi.sanjid@northsouth.edu, tanzim.hossain@northsouth.edu, shakib.junayed@northsouth.edu)}
\thanks{$^*$M. Monir Uddin is with the Department of Mathematics \& Physics, North South University, Dhaka-1229, Bangladesh (e-mail: monir.uddin@northsouth.edu)}
}



\maketitle

\begin{abstract}
Deep learning has revolutionized medical imaging by providing innovative solutions to complex healthcare challenges. Traditional models often struggle to dynamically adjust feature importance, resulting in suboptimal representation, particularly in tasks like semantic segmentation crucial for accurate structure delineation. Moreover, their static nature incurs high computational costs. To tackle these issues, we introduce Mamba-Ahnet, a novel integration of State Space Model (SSM) and Advanced Hierarchical Network (AHNet) within the MAMBA framework, specifically tailored for semantic segmentation in medical imaging.Mamba-Ahnet combines SSM's feature extraction and comprehension with AHNet's attention mechanisms and image reconstruction, aiming to enhance segmentation accuracy and robustness. By dissecting images into patches and refining feature comprehension through self-attention mechanisms, the approach significantly improves feature resolution. Integration of AHNet into the MAMBA framework further enhances segmentation performance by selectively amplifying informative regions and facilitating the learning of rich hierarchical representations. Evaluation on the Universal Lesion Segmentation dataset demonstrates superior performance compared to state-of-the-art techniques, with notable metrics such as a Dice similarity coefficient of approximately 98\% and an Intersection over Union of about 83\%. These results underscore the potential of our methodology to enhance diagnostic accuracy, treatment planning, and ultimately, patient outcomes in clinical practice. By addressing the limitations of traditional models and leveraging the power of deep learning, our approach represents a significant step forward in advancing medical imaging technology.
\end{abstract}


\section{Introduction}
\label{sec:introduction}
\IEEEPARstart{T}{he} total number of CT exams performed each year has been rising recently \cite{boland2009radiologist}, which has increased radiologists' workloads \cite{mcdonald2015effects}. Oncological radiology is anticipated to play a significant role in the anticipated increase in workload in the future, given the predicted 47\% rise in the global cancer burden from 2020 to 2040 \cite{sung2021global}. This is because patients with cancer frequently undergo multiple imaging exams over a longer period to track the progression of their disease.

% %%%%%%%%%%%%%%%%%%%%%%%%%%%%
% \begin{figure*}[htbp]

% \centering
% \arrayrulecolor{steelblue}
% \begin{tabular}{cc|cc|cc}

% \multicolumn{2}{c|}{Deeplesion} & \multicolumn{2}{c|}{Pancreas}& \multicolumn{2}{c}{Bone}\\

% Image & Mask & Image & Mask & Image & Mask \\

% \includegraphics[width=0.142\textwidth]{Figures/di_1.png} & \includegraphics[width=0.142\textwidth]{Figures/dm_1.png} & \includegraphics[width=0.142\textwidth]{Figures/pi_1.png} & \includegraphics[width=0.142\textwidth]{Figures/pm_1.png} & \includegraphics[width=0.142\textwidth]{Figures/bi_1.png} & \includegraphics[width=0.142\textwidth]{Figures/bm_1.png} \\

% \includegraphics[width=0.142\textwidth]{Figures/di_2.png} & \includegraphics[width=0.142\textwidth]{Figures/dm_2.png} & \includegraphics[width=0.142\textwidth]{Figures/pi_2.png} & \includegraphics[width=0.142\textwidth]{Figures/pm_2.png} & \includegraphics[width=0.142\textwidth]{Figures/bi_2.png} & \includegraphics[width=0.142\textwidth]{Figures/bm_2.png} \\

% \end{tabular}
% \caption{Example of cross-sectional MRI pictures and
% corresponding labels ranging from normal sizes to those occupying only a few pixels.}
% \label{fig:preprocessed_data}

% \end{figure*}
% %%%%%%%%%%%%%%%%%%%%%%%%%%%% 

%%%%%%%%%%%%%%%%%%%%%%%%%%%%
\begin{figure*}[htbp]

\centering
\arrayrulecolor{steelblue}
\begin{tabular}{cc|cc|cc}

\multicolumn{2}{c|}{Deeplesion} & \multicolumn{2}{c|}{Pancreas}& \multicolumn{2}{c}{Bone}\\

Image & Mask & Image & Mask & Image & Mask \\

\includegraphics[width=0.142\textwidth]{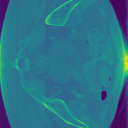} &
\begin{tikzpicture}
        % Include original image
        \node[anchor=south west,inner sep=0] (image) at (0,0) {\includegraphics[width=0.142\textwidth]{Figures/di_1.png}};
        % Include original mask with 40% transparency
        \node[anchor=south west,inner sep=0,opacity=1] (mask) at (0,0) {\includegraphics[width=0.142\textwidth]{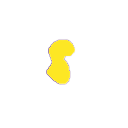}};
\end{tikzpicture} &
\includegraphics[width=0.142\textwidth]{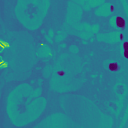} & 
\begin{tikzpicture}
        % Include original image
        \node[anchor=south west,inner sep=0] (image) at (0,0) {\includegraphics[width=0.142\textwidth]{Figures/pi_1.png}};
        % Include original mask with 40% transparency
        \node[anchor=south west,inner sep=0,opacity=1] (mask) at (0,0) {\includegraphics[width=0.142\textwidth]{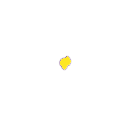}};
\end{tikzpicture} &
\includegraphics[width=0.142\textwidth]{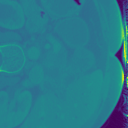} & 
\begin{tikzpicture}
        % Include original image
        \node[anchor=south west,inner sep=0] (image) at (0,0) {\includegraphics[width=0.142\textwidth]{Figures/bi_1.png}};
        % Include original mask with 40% transparency
        \node[anchor=south west,inner sep=0,opacity=1] (mask) at (0,0) {\includegraphics[width=0.142\textwidth]{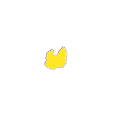}};
\end{tikzpicture} \\

\includegraphics[width=0.142\textwidth]{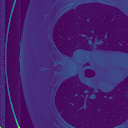} &
\begin{tikzpicture}
        % Include original image
        \node[anchor=south west,inner sep=0] (image) at (0,0) {\includegraphics[width=0.142\textwidth]{Figures/di_2.png}};
        % Include original mask with 40% transparency
        \node[anchor=south west,inner sep=0,opacity=1] (mask) at (0,0) {\includegraphics[width=0.142\textwidth]{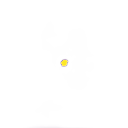}};
\end{tikzpicture} &
\includegraphics[width=0.142\textwidth]{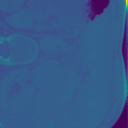} & 
\begin{tikzpicture}
        % Include original image
        \node[anchor=south west,inner sep=0] (image) at (0,0) {\includegraphics[width=0.142\textwidth]{Figures/pi_2.png}};
        % Include original mask with 40% transparency
        \node[anchor=south west,inner sep=0,opacity=1] (mask) at (0,0) {\includegraphics[width=0.142\textwidth]{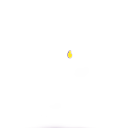}};
\end{tikzpicture} &
\includegraphics[width=0.142\textwidth]{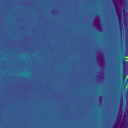} & 
\begin{tikzpicture}
        % Include original image
        \node[anchor=south west,inner sep=0] (image) at (0,0) {\includegraphics[width=0.142\textwidth]{Figures/bi_2.png}};
        % Include original mask with 40% transparency
        \node[anchor=south west,inner sep=0,opacity=1] (mask) at (0,0) {\includegraphics[width=0.142\textwidth]{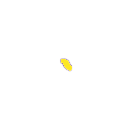}};
\end{tikzpicture} \\

\end{tabular}
\caption{Example of cross-sectional MRI pictures and
corresponding labels ranging from normal sizes to those occupying only a few pixels.}
\label{fig:preprocessed_data}

\end{figure*}
%%%%%%%%%%%%%%%%%%%%%%%%%%%% 

Assessing disease advancement and evaluating treatment efficacy in longitudinal CT scans frequently necessitate manual measurements of lesions \cite{yan2019mulan} along either the long or short axes. Typically, these measurements adhere to the Response Evaluation Criteria In Solid Tumors (RECIST) guidelines \cite{eisenhauer2009new} were formulated to streamline and standardize this procedure. These guidelines prescribe a cap on the number of lesions to be measured, restricting it to a maximum of five "target lesions" across various organs or structures through which the overall response is estimated.

% \subsection{Role of Lesion Segmentation in CAD:}
To alleviate the time constraints associated with the automated segmentation of lesions from CT scans holds significant importance across various computer-aided diagnosis (CAD) endeavors, including pathology identification \cite{peng2019self}, tumor progression tracking \cite{katzmann2022predicting} and quantitative analysis of disease advancement \cite{lu2009quantitative}. In the case of annotating lesions in oncological scans, automatic segmentation models offer a solution by extracting information with minimal input from a radiologist. 
Considerable efforts have been dedicated to the advancement of automated lesion segmentation techniques to alleviate the workload of radiologists and enhance diagnostic precision \cite{agarwal2020weakly, cai2018accurate, chunfeng2019joint, nikan2020pwd, tang2018ct, wang2017hierarchical}. This guidance may entail a single click within the lesion by a radiologist or utilizing a bounding box prediction from a detection model. Employing segmentation of the lesion volume yields supplementary data that can be exploited to calculate more comprehensive lesion volumes or characteristics. Additionally, registration algorithms can facilitate the propagation of segmented lesions \cite{hering2021whole}, leading to substantial time efficiencies during follow-up examinations.

% \subsection{Advances with CNNs and DNNs:}
In recent times, significant advancements have been achieved in the realm of AI owing to the rapid evolution of convolutional neural networks (CNNs), particularly fully convolutional networks (FCNs) \cite{long2015fully}. Significant progress has been achieved in automatic segmentation models tailored for tumors. Specifically, medical challenges centered around liver, kidney, or lung tumors have witnessed substantial enhancements in performance concerning lesion segmentation. Nevertheless, in clinical settings, a pressing demand exists for versatile and robust models capable of swiftly segmenting the myriad types of lesions found within the thorax-abdomen region.

% \subsection{Data Limitations in Medical Imaging:}
The efficacy of semantic segmentation models \cite{chen2018encoder, tian2019decoders, yang2018denseaspp, yuan2018ocnet}, underscores the pivotal role of meticulously annotated training datasets, such as Cityscapes \cite{cordts2016cityscapes} and ADE20K \cite{zhou2019semantic}, in driving performance enhancements. However, in contrast to natural images, acquiring medical images poses significant challenges due to their high privacy and confidentiality concerns. Furthermore, annotating medical images is labor-intensive and costly and demands substantial clinical expertise, resulting in a scarcity of publicly available data for segmentation tasks. Consequently, the primary obstacle in training a precise lesion segmentation model is the absence of a comprehensive dataset. 

% \subsection{Limited Diversity in Existing Datasets:}
Moreover, aside from the limited scale of available data, another issue arises from the specificity of existing medical datasets \cite{heller2019kits19, moreira2012inbreast}, which typically focus on particular lesion types such as liver, and kidney tumors \cite{heller2019kits19}, breast masses \cite{moreira2012inbreast,lee2017curated}, or lung nodules \cite{armato2011lung}.  This limitation poses a challenge in developing a comprehensive lesion segmentation framework. Consequently, prevailing lesion segmentation models often concentrate on segmenting a single lesion type from its corresponding anatomical region. However, in reality, various types of lesions exhibit interconnections. For instance, metastases can disseminate to different body parts \cite{cao2020dual, christ2017automatic, li2018h} via the lymphatic system or bloodstream.

% \subsection{Introduction of ULS Dataset:}
To address the above problems, we use The Universal Lesion Segmentation '23 Challenge dataset. The dataset consists of 6,514 fully 3D annotated lesions of multiple body parts. Although prior efforts have been devoted to ULS \cite{cai2018accurate, tang2020one, tang2021lesion} development, most research in this domain has heavily relied upon a single partially annotated dataset \cite{yan2018deeplesion}, which solely encompasses the long-and short-axis diameters on a solitary axial slice. 

% \subsection{Challenges in Universal Lesion Segmentation:}
Lesions of various types typically manifest diverse characteristics in size, shape, and appearance. For instance, depicted in Figure \ref{fig:preprocessed_data} are lesions ranging from normal sizes to those occupying only a few pixels. Additionally, a multitude of shapes and appearances of lesions are observable. Consequently, employing existing segmentation techniques directly \cite{oktay2018attention, ronneberger2015u, zhang2018context} for universal lesion segmentation is deemed suboptimal. 
Numerous initiatives have emerged to automate the process of measuring lesion size. Notably, deep convolutional neural networks have been effectively deployed to delineate tumors within various anatomical regions such as the brain \cite{havaei2017brain}, lungs \cite{jin2018ct, wang2017central}, pancreas \cite{zhang2020robust}, liver \cite{christ2017automatic, chlebus2018automatic} and enlarged lymph nodes \cite{nogues2016automatic, zhu2020lymph}. While many of these methodologies are tailored to specific lesion types, the ideal lesion size measurement tool should be able to manage a diverse array of lesions encountered in clinical practice.

Recently, several deep learning (DL) algorithms aimed at universal lesion detection (ULD) or segmentation (ULS) in CT images have demonstrated notable success in aiding the diagnosis of various medical conditions \cite{yan2019mulan, lyu2021segmentation, tang2021weakly, yan2020learning}. These investigations have revealed that ULS algorithms, drawing from diverse datasets encompassing various organs, exhibit enhanced efficacy and scalability compared to those tailored for singular organ lesion segmentation. Furthermore, in response to the challenges posed by the laborious and costly nature of manual image annotations in medical contexts, ULS algorithms trained on diverse ultrasound (US) image datasets spanning multiple organs have shown promise in enhancing the quality of annotations in novel datasets through transfer learning. This approach effectively mitigates the burdens associated with the demanding requirements for both quantity and quality of medical image annotations.

% \subsection{Introduction of AHNet \& Integration of Mamba and Ahnet}
To address the problems discussed earlier and to design a better universal lesion segmentation model, we propose a unique methodology that merges to address semantic universal lesion segmentation tasks. The methodology incorporates the Advanced Hierarchical Network (AHNet) model into the selective state space model (MAMBA) framework \cite{gu2023mamba}, inspired by the integration of HUNet into MAMBA as described by \cite{sanjid2024integrating}, incorporating attention mechanisms and residual blocks to enhance semantic segmentation performance. Firstly, the AHNet model, a customized
UNet architecture is seamlessly integrated into the MAMBA architecture, extending its capabilities in medical image segmentation. The integration process between the state space model Mamba and the AHNet architecture occurred by incorporating
Mamba blocks alongside attention gates and residual blocks within the AHNet model’s upsampling pathway.

% \subsection{Overall contribution list:} 
To conduct a thorough comparative analysis between our methodology and previous approaches, we have constructed an exhaustive benchmark encompassing established medical and semantic image segmentation techniques. Our evaluation employs four widely recognized metrics in medical image segmentation. The experimental findings affirm that our approach exhibits superior performance compared to prior state-of-the-art methods, thereby establishing itself as a robust benchmark for future research endeavors. This comprehensive benchmark serves as a valuable resource for future investigations and underscores our contributions.
We summarize our contributions as follows:

\begin{itemize}
    \item HUNet in \cite{sanjid2023prostate} is enhanced by AHNet with upsampling attention and residual blocks, resulting in improved segmentation accuracy in medical imaging and computer vision tasks.
    \item The attention-enhanced upsampling block, derived from Hunet, is introduced in AHNet, enabling dynamic feature importance adjustment and boosting semantic segmentation accuracy, surpassing standard upsampling blocks in segmentation tasks with enhanced feature representation and semantic information propagation.
    \item The ability of AHNet to capture intricate patterns is further enhanced by integrating Mamba blocks, leading to enhanced segmentation performance in biomedical image analysis.
    \item Input features are selectively emphasized by upsampling attention gates, enhancing reconstruction fidelity in tasks such as image segmentation.
    \item Fine details are preserved during reconstruction by upsampling residual blocks, contributing to enhanced performance in image segmentation tasks.
    \item The superior performance of Mamba-AHNet with image reconstruction is highlighted by achieving the best result on the ULS23 dataset. This outcome is further clarified and supported in Section \ref{Results}.
    \item Segmentation accuracy is improved by integrating upsampling image reconstruction with upsampling attention mechanisms, emphasizing relevant image regions.
    \item Developed a user-friendly website with API integration that facilitates efficient segmentation using the Mamba-AHNet with image reconstruction, ensuring quick analysis turnaround times in clinical settings for improved diagnosis and treatment planning.
\end{itemize}

The rest of the paper is structured into several sections and subsections to provide a clear presentation of the research. It begins with an examination of related works (Section~\ref{Related Works}), followed by an exploration of preliminaries (Section ~\ref{Preliminaries}) and a detailed explanation of the methodology used (Section~\ref{Methodology}). The experiments conducted are outlined in Section~\ref{Experiments}, with specific attention given to the dataset (Subsection~\ref{Dataset}), data preprocessing (Subsection~\ref{Data_Preprocessing}), and evaluation metrics (Subsection~\ref{Evaluation_Metrics}). Computational aspects, including computation time, are discussed in Section ~\ref{Computation_Time}. The numerical analysis of results is presented in Section~\ref{Numerical_Analysis}, with subsections dedicated to presenting the outcomes (Subsection~\ref{Results}) and discussing their implications (Subsection~\ref{Discussion}). The paper concludes with a summary of findings in Section~\ref{Conclusion}, followed by acknowledgment of each author's contributions in Section~\ref{Authors Contribution}.

\section{Related Works}
\label{Related Works}
Recently, CNNs, the Vision-language model, and Attention mechanisms have been used for medical image segmentation. Most of these methodologies are tailored to specific lesion types.

Semantic segmentation can be understood as classifying each pixel in an image, akin to pixel-level image classification. Consequently, numerous image classification networks have been extended to facilitate semantic segmentation \cite{long2015fully, chen2017deeplab, xu2022mrdff, xu2021multi}. Notably, FCN is widely recognized as the initial end-to-end pixel-to-pixel network for semantic segmentation. Among these approaches, U-Net stands out as a seminal contribution to medical image segmentation \cite{ronneberger2015u}. Building upon this foundation, UNet++ \cite{zhou2018unet++} has enhanced the skip connection mechanism of U-Net. Nonetheless, many of these methods exhibit sensitivity to the data's quantity and quality, thereby constraining their generalization performance. To address this limitation, several approaches have investigated the utility of semi-supervised learning across diverse domains \cite{liu2020semi, li2022semi, xia2020uncertainty}. Moreover, the challenge posed by insufficient data and annotations can be alleviated by incorporating multiple modalities into the learning models.

CLIP \cite{radford2021learning} stands as a groundbreaking venture into the realm of large-scale Vision-Language Pretraining (VLP) models. It employs contrastive learning to imbibe image representations from a vast dataset comprising 400 million pairs of images and corresponding text descriptions. To streamline visual input processing, ViLT \cite{kim2021vilt} capitalizes on interaction layers to handle visual features without necessitating a separate deep visual embedder. Consequently, many studies have emerged, focusing on enhancing image segmentation by integrating textual information. For instance, Ding et al. \cite{ding2022vlt} devised a Vision-Language Transformer (VLT) framework tailored for referring segmentation tasks. This framework fosters deep interactions among multimodal data streams, harmonizing linguistic and visual features. In contrast, the Language-Aware Vision Transformer (LAVT) framework \cite{yang2022lavt} employs an early fusion approach, leveraging a pixel-word attention mechanism to blend linguistic and visual features seamlessly. While the influence of VLP extends to medical image analysis \cite{bhalodia2021improving, muller2022joint, tomar2022tganet}, the unique characteristics of medical images pose distinct challenges. Unlike natural images, medical images often exhibit blurred boundaries between different regions, rendering precise segmentation formidable. Consequently, the direct application of vision-language models to medical image analysis proves impractical. VLT and LAVT, tailored for reference segmentation in natural images, rely on explicit alignment between image and text, a feat difficult to achieve in the context of medical images due to inherent disparities. Similarly, ViLT, geared towards multimodal tasks like Visual Question Answering (VQA), may not seamlessly translate to the nuances of medical image analysis.

Beginning with the advent of attention mechanisms in computer vision as evidenced by RAN \cite{wang2017residual}, researchers have made strides in integrating them into this domain. Authors in  \cite{woo2018cbam} introduced the widely recognized Convolutional Block Attention Module (CBAM), which employs spatial and channel attention to refine features dynamically. Furthermore, alongside traditional attention mechanisms \cite{huang2020deep}, self-attention mechanisms \cite{vaswani2017attention} have made their foray into computer vision. Nevertheless, since self-attention was initially tailored for addressing Natural Language Processing (NLP) tasks, it confronts several hurdles, notably elevated computational overhead and the oversight of local image features. 

\cite{lyu2021segmentation} propose a novel model integrating segmentation assistance alongside conventional detection, employing a mutual-mining strategy between the two branches to jointly identify lesions and mitigate adverse effects through selective loss and exclusion of suspicious lesions from fine-tuning. \cite{yan2019mulan} present an end-to-end framework structured around an enhanced Mask RCNN, incorporating detection, tagging, and segmentation branches. Additionally, \cite{wu2023uls4us} develop a universal lesion detection algorithm, starting with a framework that learns from diverse lesion datasets concurrently and utilizes proposal fusion, followed by methodologies to extract absent annotations from datasets with incomplete labeling through clinical prior knowledge and knowledge transfer across datasets. These advancements collectively contribute to more effective lesion detection and segmentation in medical imaging.

\section{Preliminaries}
\label{Preliminaries}
\subsection{State Space Models}
The SSMs are a class of systems that relate a one-dimensional function or sequence $u(t) \longmapsto y(t) \in \mathcal{R}$ and can be expressed as a linear time-invariant (LTI) continuous-time system of the form \ref{sss}
\begin{equation}
\begin{aligned}
  \dot{x}(t) &= \textbf{A}x(t) + \textbf{B}u(t), \\
  y(t) &=  \textbf{C}x(t) +\textbf{D}u(t),
\end{aligned}
\label{sss}
\end{equation}

where $\textbf{A} \in \mathbb{R}^{n \times n}$ and $\textbf{B}\in\mathbb{R}^{n\times 1}$ and $\textbf{C} \in \mathbb{R}^{1\times n}$, are representing the system, input and output matrices respectively, of the system. The vectors  $x(t)$, $u(t)$, and $y(t)$, respectively, are known as the state, input, and output of the system.  These models possess advantageous features like linear computational complexity per time step and parallelized computation, facilitating efficient training. Conventional SSMs require more memory than equivalent convolutional neural networks (CNNs). They frequently face challenges such as vanishing gradients during training, limiting their broader utility in general sequence modeling. 

The S4 \cite{gu2021efficiently} enhances conventional SSMs by introducing structured patterns in the state matrix $A$ and employing a proficient algorithm. In particular, the state matrix is constructed and initialized using the High-Order Polynomial Projection Operator (HIPPO) \cite{gu2020hippo}, creating deep sequence models with enhanced capabilities for efficient long-range reasoning. S4, as an innovative network architecture, has outperformed Transformers \cite{vaswani2017attention} by a considerable margin in the demanding Long Range Arena Benchmark \cite{tay2020long}. Mamba introduces notable advancements in discrete data modeling using SSMs, particularly in domains such as text and genomics. Two primary enhancements distinguish Mamba from traditional SSMs approaches. Firstly, it incorporates an input-dependent selection mechanism, unlike conventional SSMs, invariant to time and input, enabling efficient information extraction tailored to specific inputs. This is achieved by parameterizing the SSMs parameters based on the input data. Secondly, Mamba introduces a hardware-aware algorithm that scales linearly with sequence length, facilitating recurrent model computation through scanning, thereby outperforming previous methods on modern hardware. The Mamba architecture integrates SSM blocks with linear layers, is notably simpler, and has exhibited state-of-the-art performance across various long-sequence domains such as language and genomics. This underscores its significant computational efficiency in both the training and inference stages.

\subsection{Discrete-time SSM: The Recurrent Representation}
%%%%%%%%%%%%%%%%%%%%%%%%%%%%%%%%%%%%%%%%%%
In discretizing \ref{sss}, the granularity of the input is indicated by the step size \textbf{$\Delta$}. This is required when working with a discrete input sequence $(u_0, u_1, \ldots)$ rather than a continuous function $u(t)$. As $u_k = u(k\Delta)$, the inputs are essentially samples of an underlying continuous signal $u(t)$. 

As demonstrated in earlier research, including \cite{tustin1947method}, we use the bilinear approach to discretize the continuous-time State Space Model (SSM). The state matrix $\textbf{A}$ is converted into an approximation $\mathbf{\overline{A}}$ using this procedure. The discrete SSM is

\begin{equation}
\begin{aligned}
  \mathbf{\overline{A}} &= (\mathbf{\overline{I}} - \Delta/2.\mathbf{\overline{A}})^{-1}(\mathbf{\overline{I}} + \Delta/2.\mathbf{\overline{A}})\\
  \mathbf{\overline{B}} &= (\mathbf{\overline{I}} - \Delta/2.\mathbf{\overline{A}})^{-1}\Delta \mathbf{\overline{B}}\\
  \mathbf{\overline{C}} &= \mathbf{\overline{C}}\\
  y_k &= \mathbf{\overline{C}}x_k\\
  x_k &= \mathbf{\overline{A}}x_k + \mathbf{\overline{B}}u_k
\end{aligned}
\label{fds}
\end{equation}

Rather than treating $u_k$ in \ref{fds} as a function that maps from sequence-to-sequence, we now consider it as a sequence-to-sequence mapping to $y_k$. The discrete State-Space Model (SSM) may now be computed similarly to a Recurrent Neural Network (RNN) because the state equation also recurs in $x_k$. To be more precise, a hidden state $x_k \in \mathbb{R}^N$ has a transition matrix $\mathbf{\overline{A}}$.

In this study, the discretized SSM matrices are denoted as $\mathbf{\overline{A}}, \mathbf{\overline{B}}, \ldots$, which are found using \ref{fds}. It is important to remember that these matrices depend on a step size $\Delta$ in addition to $\textbf{A}$. When this reliance is acknowledged, we leave it out of our notation explicitly for the sake of concision.

\begin{figure*}[t]
     \centering
     \includegraphics[width=\textwidth]{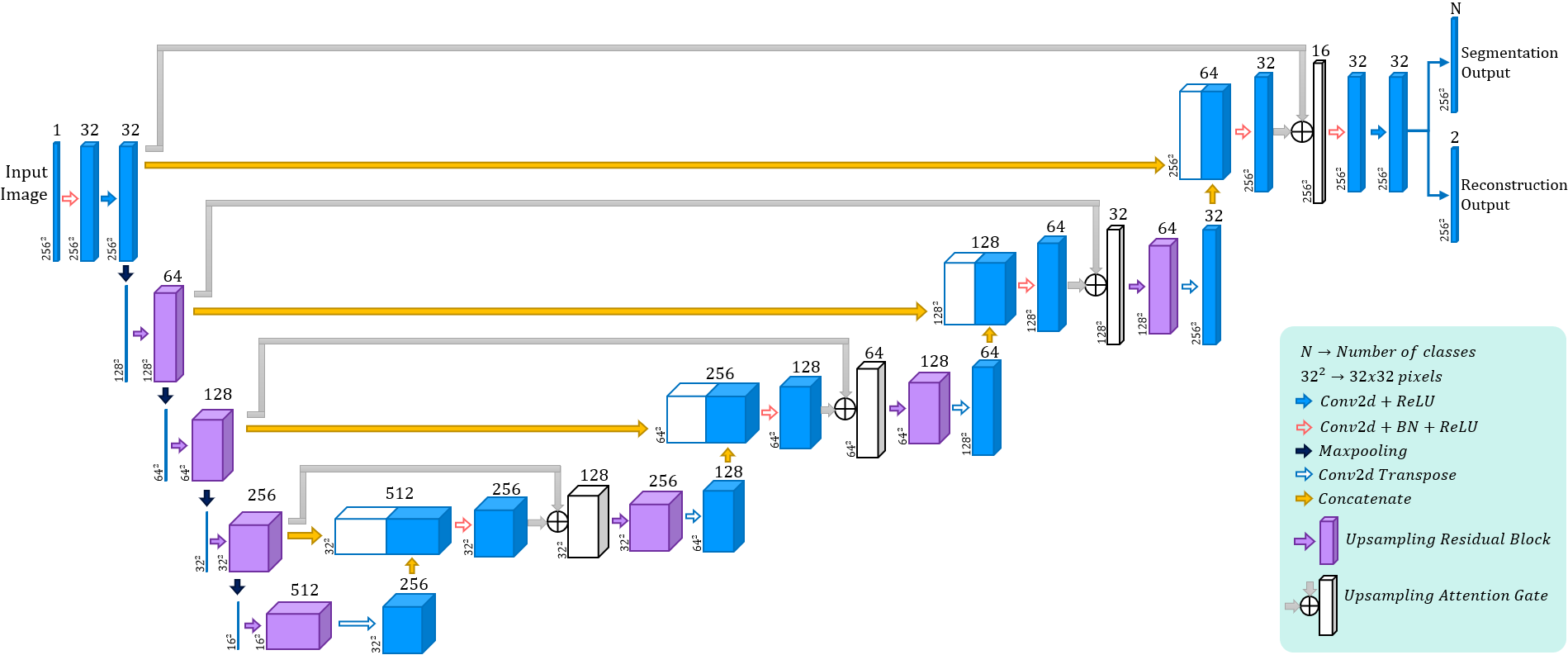}
     \caption{An illustration of AHNet architecture}
     \label{fig:ahnet}
\end{figure*}

\subsection{Training SSMs: The Convolutional Representation}

Since the recurrent SSM is sequential, training on modern hardware presents practical issues \ref{fds}. Yet, a well-established connection has been identified between continuous convolutions and linear time-invariant (LTI) SSMs such as \ref{sss}. This allows for the rewriting of \ref{fds} as a discrete convolution. 

To simplify, assume the initial state is $x_{-1} = 0$. Then unrolling \ref{fds} explicitly yields

\begin{align*}
  x_0 &= \mathbf{\overline{B}}u_0 \\
  x_1 &= \mathbf{\overline{AB}}u_0 + \mathbf{\overline{B}}u_1 \\
  x_2 &= \mathbf{\overline{A}}^2\mathbf{\overline{B}}u_0 +\mathbf{\overline{AB}}u_1 + \mathbf{\overline{B}}u_2 \\ 
  \vdots\\
   y_0 &= \mathbf{\overline{CB}}u_0\\
   y_1 &= \mathbf{\overline{CAB}}u_0 + \mathbf{\overline{CB}}u_1\\
   y_2 &= \mathbf{\overline{CA}}^2\mathbf{\overline{B}}u_0 + \mathbf{\overline{CAB}}u_1 + \mathbf{\overline{CB}}u_2\\
   \vdots
\end{align*}

This can be vectorized into a convolution \ref{jytj} with an explicit formula for the convolution kernel \ref{fghdf}.

\begin{equation}
\begin{aligned}
  y_k &= \mathbf{\overline{CA}}^k\mathbf{\overline{B}}u_0 + \mathbf{\overline{CA}}^{k-1}\mathbf{\overline{B}}u_1 + \hdots + \mathbf{\overline{CAB}}u_{k-1} + \mathbf{\overline{CB}}u_k,\\
  y &= \mathbf{\overline{K}} * u.
\end{aligned}
\label{jytj}
\end{equation}

\begin{equation}
\begin{aligned}
  \mathbf{\overline{K}}\in \mathbb{R}^L &:= \mathcal{K}_L(\mathbf{\overline{A}}, \mathbf{\overline{B}}, \mathbf{\overline{C}})\\ &:= (\mathbf{\overline{CA}}^i\mathbf{\overline{B}})_{i \in [L]} \\&= (\mathbf{\overline{CB}}, \mathbf{\overline{CAB}} \hdots, \mathbf{\overline{CA}}^{L-1}\mathbf{\overline{B}}).
\end{aligned}
\label{fghdf}
\end{equation}
Put otherwise, given $\mathbf{\overline{K}}$, \ref{jytj} indicates a solitary (non-circular) convolution that can be computed well with FFTs. Nonetheless, $\mathbf{\overline{K}}$ determination in \ref{fghdf} is intricate and serves as the foundation for our technological advances in Section \ref{Methodology}. We call the \textbf{SSM convolution kernel} or filter $\mathbf{\overline{K}}$.

\subsection{Vision Mamba Layer (VM Layer):}
The original SSM block is improved by the VM Layer for the extraction of deep semantic features from images. In particular, it nearly eliminates the need for new parameter introduction and computing complexity by leveraging sophisticated residual connections and adjustment factors to improve the long-range spatial modeling capabilities of SSM significantly. Given the input deep feature $M_{in}^l \in \mathbb{R}^{L \times C}$, as shown in Figure \ref{fig:mamba_ahnet}{\color{cite-blue}(a)}, the VM Layer first uses LayerNorm and then the Vision State Space Model (VSSM) to capture spatial long-range dependencies. For improved performance, it then uses an adjustment factor $s \in \mathbb{R}^C$ in the residual connection \cite{chen2023recursive}. This process can be represented mathematically as follows:

\begin{equation}
    \widetilde{M^l} = VSSM(L_{norm}(M_{in}^l)) + s.M_{in}^l
\end{equation}

where $L_{norm}$ is the Layer Norm. The RVM Layer then uses a second LayerNorm to normalize $\widetilde{M^l}$. Finally, it uses a projection layer to transform $\widetilde{M^l}$ into a more profound feature. The above process can be formulated as:

\begin{equation}
    M_{out}^l = Projection(L_{norm}(\widetilde{M^l}))
\end{equation}

\subsection{Vision State-Space Module (VSS Module):}
\begin{figure*}[htb!]
     \centering
     \includegraphics[width=\textwidth]{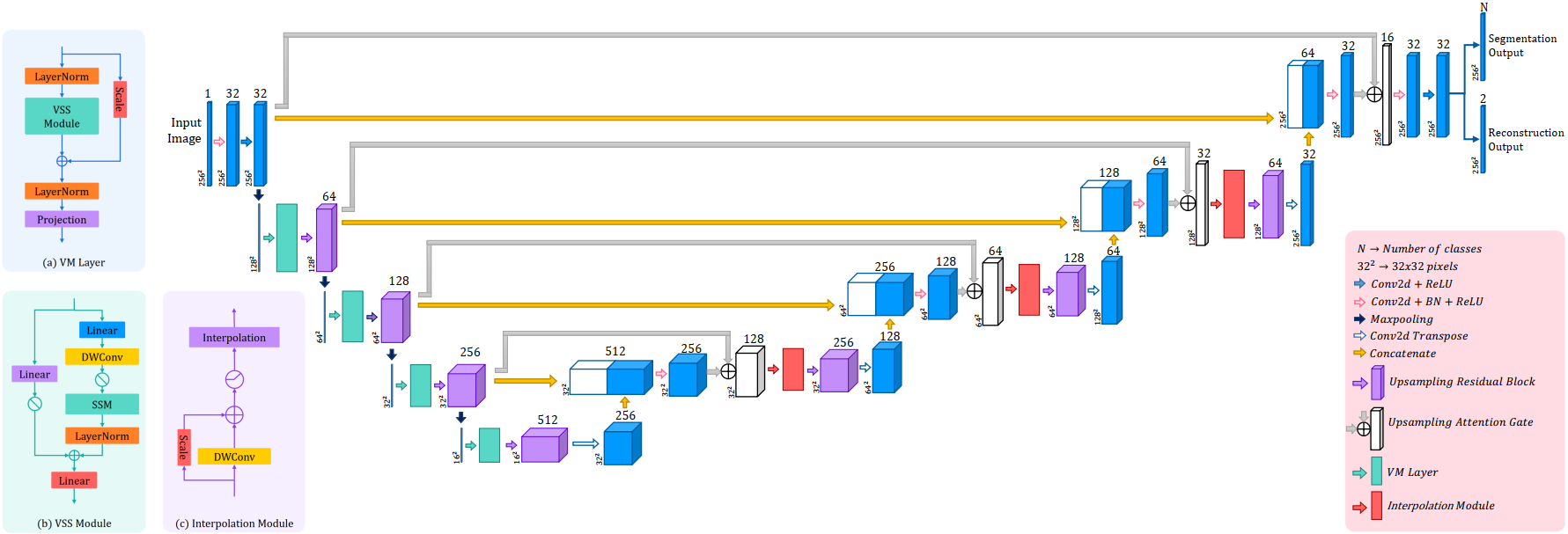}
     \caption{An illustration of Mamba-AHNet with image reconstruction architecture}
     \label{fig:mamba_ahnet}
\end{figure*}

For long-range spatial modeling, \ref{fig:mamba_ahnet}{\color{cite-blue}(b)} is comparable to the methodology described in \cite{liu2024vmamba}. The feature $W_{in}^l \in \mathbb{R}^{L\times C}$ is fed into two parallel branches by the VSS Module. The VSS Module uses a linear layer in the first branch to expand the feature channels to $\lambda \times C$, where $\lambda$ is a pre-defined channel expansion factor. It then applies the SSM and LayerNorm, then a DWConv and SiLU activation function \cite{shazeer2020glu}. The SiLU activation function is applied after the VSS Module uses a linear layer to increase the feature channels $\lambda \times C$ in the second branch. The output $W_{out}$ has the same form as the input $W_{in}$ because the VSS Module then uses the Hadamard product to combine the features from the two branches and projects the channel number back to $C$. The above process can be formulated as: 

\begin{equation}
\begin{aligned}
  W_1 &= L_{norm}(SSM(SiLU(DW Conv(Linear(W_{in})))))\\
  W_2 &= SiLU(Linear(W_{in}))\\
  W_{out} &= Linear(W_1 \odot W_2)
\end{aligned}
\label{jytj}
\end{equation}

\section{Methodology}
\label{Methodology}

Our methodology merges the innovative  SSM with the CNN-based architecture, integrated into the MAMBA framework to address semantic segmentation tasks. The VSSM methodology begins with dissecting images into patches, then progressing through VSSLayer encoder and decoder modules. Within these layers, the SS2D mechanism orchestrates self-attention, complemented by residual connections and stochastic depth regularization, refining the model's comprehension. This iterative process of compression and expansion enhances the feature resolution significantly.

Simultaneously, the methodology incorporates the AHNet model into the MAMBA framework, incorporating attention mechanisms and residual blocks to enhance semantic segmentation performance. Firstly, the AHNet model, a customized HUNet architecture \cite{sanjid2023prostate}, is seamlessly integrated into the MAMBA architecture, extending its capabilities in medical image segmentation.

The integration involves adapting the AHNet model's architecture to fit the MAMBA framework while preserving unique features such as attention mechanisms and residual blocks. Attention mechanisms selectively amplify informative regions in feature maps by computing attention weights based on the interaction between feature maps. This mechanism enhances the model's ability to focus on salient features while suppressing irrelevant information, improving segmentation accuracy. Additionally, residual blocks play a crucial role in feature extraction and propagation. By introducing skip connections and residual learning, these blocks facilitate the learning of rich hierarchical representations, enabling effective feature propagation across different scales and depths within the network architecture. This adaptive adjustment further enhances the model's capacity to capture and propagate features, improving segmentation performance. Combining the AHNet model with attention mechanisms and residual blocks within the MAMBA framework enhances semantic segmentation accuracy. It contributes to the model's interpretability, robustness, and scalability across diverse medical imaging datasets and segmentation tasks.

\subsection{Ahnet:} The AHNet model in Figure \ref{fig:ahnet} introduces several innovative features to enhance feature importance adjustment and learning capabilities. Feature Importance Adjustment in the context of AHNet refers to the dynamic modulation of the relevance or importance of different features extracted from medical images during the segmentation process. Firstly, it incorporates upsampling attention blocks, which dynamically adjust the importance of features at different spatial locations. This attention mechanism is achieved by computing the attention map through convolutional operations between the input and corresponding features from the contracting path. This allows the model to focus on relevant features while discarding irrelevant ones, enhancing feature representation. The model uses an upsample block, which concatenates the upsampled feature maps with the feature maps from the contracting path to provide fine-grained details and spatial context. Additionally, the model utilizes upsampling residual blocks to facilitate learning complex feature representations through residual connections and multiple convolutional layers. These residual blocks improve gradient flow and training stability and enable the model to capture intricate patterns and details in the data. Furthermore, the model employs standard convolutional and pooling layers for downsampling followed by transposed convolutional layers for upsampling, ensuring efficient feature extraction and reconstruction. The AHNet model improves upon the traditional UNet architecture by addressing its limitations. While UNet has been widely used and proven effective for semantic segmentation tasks, it lacks explicit mechanisms to focus on relevant features and capture intricate details in the data. Traditional UNet architectures rely solely on skip connections and transposed convolutions for upsampling. This may lead to the loss of spatial information and inadequate feature representation, especially in regions with complex structures or subtle variations. Additionally, UNet does not incorporate attention mechanisms to dynamically adjust feature importance based on spatial context, potentially resulting in suboptimal segmentation performance, particularly in scenarios with significant variances in object sizes or shapes. By integrating upsampling attention blocks and residual blocks, the AHNet model overcomes these limitations by enhancing feature extraction, capturing fine details, and improving segmentation accuracy, making it a more robust and versatile architecture for various medical imaging and computer vision tasks.

\subsection{Intregating Mamba with Ahnet:}
The integration process between the state space model Mamba and the AHNet architecture occurred by incorporating Mamba blocks alongside attention gates and residual blocks within the AHNet model's upsampling pathway. Initially, the transformation parameters of the SSM were manipulated to ensure efficient computation and modeling of long-range dependencies. These matrices underwent discretization to convert continuous parameters to discrete ones, facilitating resolution invariance and proper model normalization. Subsequently, within the AHNet architecture, the upsampling blocks were augmented with Mamba blocks, attention gates, and residual blocks to enhance the model's ability to capture intricate patterns and relationships within the feature maps. This integration process enabled the AHNet model to effectively recover spatial details while simultaneously capturing complex patterns and relationships across different spatial scales, leading to improved segmentation performance in biomedical image analysis tasks. The architecture is shown in Figure \ref{fig:mamba_ahnet}.
\begin{algorithm}[t]
	\SetAlgoVlined
	\KwIn{Input tensor $x$, Skip connection tensor $t$, Number of filters $f$, Kernel size $k$, Padding method $p$ and Strides $s$.}
    \KwOut{Tensor $x_{up}$.}
	\caption{Attention-Enhanced Upsampling Block} 
	\label{alg:attention_enhanced_upsampling_block}

    Upsample the input tensor $x$ using transpose convolution with $f$, $k$, $p$, and $s$\\
    Apply batch normalization to the upsampled tensor $x_{up}$\\
    Apply ReLU activation function element-wise to the normalized tensor $x_{up}$\\
    Compute an attention map $A(x, g, d)$ by applying an attention mechanism to $x_{up}$, the skip connection tensor $t$, and $\frac{filters}{2}$\\
    Concatenate the upsampled tensor $x_{up}$ with the skip connection tensor\\
    Apply a convolutional layer to the concatenated tensor, using $f$, $k$, and $p$, and ReLU activation\\
    Apply batch normalization to the resulting tensor\\
    Apply ReLU activation function element-wise to the normalized tensor\\
    Return the resulting tensor $x_{up}$ as the output of the upsampling block\\
\end{algorithm}

\begin{algorithm}[t]
	\SetAlgoVlined
	\KwIn{Input tensor $x$, Number of filters $f$, Kernel size $k$, Padding method $p$ and Strides $s$.}
    \KwOut{Residual tensor $r$.}
	\caption{Upsampling Residual Block Function} 
	\label{alg:residual_block_function}
    Upsample the input tensor $x$ using transpose convolution with $f$, $k$, $p$, and $s$\\
    Apply batch normalization to the upsampled tensor $x$\\
    Apply ReLU activation function element-wise to the normalized tensor $x$\\
    Compute an attention map $A(x, g)$ by applying an attention mechanism to $x$, the skip connection tensor, and $\frac{f}{2}$\\
    Concatenate the upsampled tensor $x$ with the skip connection tensor\\
    Apply a convolutional layer to the concatenated tensor, using $f$, $k$, and $p$, and ReLU activation\\
    Apply batch normalization to the resulting tensor\\
    Apply ReLU activation function element-wise to the normalized tensor\\
    Return the resulting tensor as the output of the upsampling block\\
\end{algorithm}

\begin{algorithm}[t]
	\SetAlgoVlined
	\KwIn{Input tensors $x$, $g$, Number of channels $c$, Attended feature tensor $\varrho$.}
    \KwOut{Output tensor $att_x$.}
	\caption{Upsampling Attention Gate} 
	\label{alg:upsample_attention_gate}
	%
       % $g_{\text{cropped}} \rightarrow g[:, \text{offset\_height} : \text{offset\_height} + x\_height, \text{offset\_width} : \text{offset\_width} + x\_width, :]$

    Calculate Shape of $x$ and $g$\\
    Calculate Offset for Height\\
    Calculate Offset for Width\\
    Crop $g$ to Match $x$ Shape as $g_{cropped}$\\
    Theta Transformation:
    \quad $\theta_x \gets \text{Conv2D}(x, c, (1, 1), (1, 1), \text{padding='same'})$\\
    Batch Normalize $\theta_x$\\
    Phi Transformation:\\
    \quad $\phi_g \gets \text{Conv2D}(g_{cropped}, c, (1, 1), (1, 1), \text{padding='same'})$\\
    Batch Normalize $\phi_g$\\
    Calculate Attention Map:\\
    \quad $f \gets \text{ReLU}(\theta_x + \phi_g)$
    $\psi_f \gets \text{Conv2D}(f, 1, (1, 1), (1, 1), \text{padding='same'})$\\
    Batch Normalize $\psi_f$\\
    $rate \gets \text{Sigmoid}(\psi_f)$\\
    Apply Attention to Input Features:\\
    \quad $att_x \gets x \odot rate$\\
    \textbf{return} Attended Features $att_x$
\end{algorithm}

\subsection{Attention-Enhanced Upsampling Block}
The attention-enhanced upsampling block described in Algorithm \ref{alg:attention_enhanced_upsampling_block} is a modification from the upsampling block in HUNet, integrates attention mechanisms to bolster feature selection during the expansive path of the network. This modification becomes imperative for image segmentation tasks because it can selectively focus on relevant spatial locations from the contracting path. By dynamically adjusting the importance of different features based on their contextual relevance, the attention mechanism enables the model to capture intricate details crucial for accurate segmentation. Additionally, integrating attention mechanisms enhances the semantic information flow between different resolutions, facilitating the propagation of essential features throughout the network. Overall, this modification aims to improve feature representation and selection capabilities, thereby enhancing the segmentation performance of the model and making it more adept at handling various complexities inherent in image segmentation tasks.

The attention-enhanced upsampling block significantly improves HUnet's upsampling block by incorporating an attention mechanism. This mechanism allows the model to selectively focus on relevant spatial locations from the contracting path, dynamically adjusting the importance of different features based on their contextual relevance. The upsampling block facilitates enhanced feature selection by integrating attention, ensuring the model captures intricate details crucial for accurate segmentation. Furthermore, the attention mechanism enhances the semantic information flow between different resolutions, enabling better propagation of essential features throughout the network. This improves feature representation and selection capabilities, leading to superior segmentation performance. In contrast, the standard upsampling block cannot selectively attend to informative regions, potentially limiting its ability to capture fine details and contextual information essential for accurate segmentation. Thus, the attention-enhanced upsampling block represents a significant advancement in semantic segmentation architectures, offering improved performance and robustness in handling various complexities inherent in image segmentation tasks.

\subsection{Upsampling Residual Block Function:} The upsampling residual block function described in Algorithm \ref{alg:residual_block_function} is designed to be employed in the decoder (upsampling) portion of a neural network architecture where it plays a critical role in reconstructing high-resolution feature maps. This block includes convolutional layers followed by batch normalization, activations, residual connections, upsampling operations, and concatenation with skip connections from the encoder. Integrating skip connections effectively leverages the rich spatial information captured in the encoder, allowing for the recovery of fine details lost during downsampling. The residual connections aid in the propagation of features and gradients, fostering stable and efficient training. Consequently, the upsampling residual block contributes to the preservation of spatial details and the overall performance enhancement in tasks such as image segmentation or image generation.

In essence, the upsampling residual block outperforms a typical residual block in scenarios where the reconstruction of high-resolution feature maps is crucial. Its incorporation of upsampling operations and utilization of skip connections make it particularly effective in tasks that demand the preservation of fine details and spatial information, ultimately leading to superior performance in tasks like image segmentation and image generation.
% {\color{red}
% \begin{algorithm}[t]
% \caption{Reconstruction Output Branch:}

% \textbf{Input:} Output feature map from the last convolutional layer of the contraction path, denoted as $conv_{\text{last}} \in \mathbb{R}^{H \times W \times C}$, where $H$ is the height, $W$ is the width, and $C$ is the number of channels.

% \textbf{Convolutional Layer Operation:}
% \begin{align*}
% Reconstruction\_Output &= \text{Conv2D}(conv_{\text{last}}, n\_classes, (3, 3), \text{softmax}, \text{Same}) \\
%     &= \sigma\left(W_{\text{conv}} * conv_{\text{last}} + b_{\text{conv}}\right)
% \end{align*}
% where $\sigma$ denotes the Softmax activation function, $W_{\text{conv}}$ is the set of convolutional filters, $*$ denotes the convolution operation, and $b_{\text{conv}}$ is the bias term.

% \textbf{Output:} Reconstructed image tensor, denoted as $reconstruction\_output \in \mathbb{R}^{H \times W \times n\_classes}$.
% \end{algorithm}

\begin{algorithm}[t]
	\SetAlgoVlined
	\KwIn{Output feature map from the last convolutional layer of the contraction path, denoted as $conv_{\text{last}} \in \mathbb{R}^{H \times W \times C}$, where $H$ is the height, $W$ is the width, and $C$ is the number of channels.}
    \KwOut{Reconstructed image tensor, denoted as $re_{out} \in \mathbb{R}^{H \times W \times N}$ where $N$ is the no. of classes.}
	\caption{Reconstruction Output Branch} 
	\label{alg:reconstruction_output_branch}
 \textbf{Convolutional Layer Operation:}
\begin{align*}
re_{out} &= \text{Conv2D}(conv_{\text{last}}, N, (3, 3), \text{softmax}, \text{Same}) \\
    &= \sigma\left(W_{\text{conv}} * conv_{\text{last}} + b_{\text{conv}}\right)
\end{align*}
where $\sigma$ denotes the Softmax activation function, $W_{\text{conv}}$ is the set of convolutional filters, $*$ denotes the convolution operation, and $b_{\text{conv}}$ is the bias term.
The "Same" padding option ensures that the output feature map has the exact spatial dimensions as the input feature map. Padding with zeros is applied to the input feature map to achieve this, preserving the spatial information during convolution.
    
\end{algorithm}

% \textbf{Reconstruction Output Branch:}

% \textbf{Input:} Output from the last convolutional layer of the contraction path

% \begin{enumerate}
%     \item Apply a convolutional layer to capture spatial information and learn features:
%     \begin{itemize}
%         \item Input: Output from the last convolutional layer
%         \item Output: $conv\_output$
%         \item Filters: $n\_classes$ (number of channels in the input images)
%         \item Kernel Size: $(3, 3)$
%         \item Activation Function: Sofmax
%         \item Padding: Same
%     \end{itemize}

%     \item \textbf{Output:} $reconstruction\_output$ (Resultant tensor representing the reconstructed image)
% \end{enumerate}
% Please Check Right Side also color is not working on algorithm 4

\subsection{Upsampling Image reconstruction Branch:}
The reconstruction branch in Ahnet first receives feature maps from different scales of the encoder-decoder architecture used for segmentation. These feature maps undergo an attention mechanism, fusing with corresponding features to focus on relevant regions. This attention-enhanced information is then propagated through residual blocks to extract hierarchical representations. The reconstruction branch integrates skip connections to preserve detailed information from the encoder. After passing through convolutional layers, the reconstructed output is generated to faithfully recreate the input image while emphasizing regions crucial for segmentation. During training, the reconstruction branch is jointly optimized with the segmentation task, incorporating a combined loss function to balance both objectives and improve overall model performance effectively.

The upsampling image reconstruction branch in the Ahnet differs from a standalone reconstruction model's tailored integration within the segmentation framework. Unlike a typical reconstruction model, which focuses solely on faithfully reproducing the input image, the reconstruction branch here is designed to complement the segmentation task by emphasizing relevant regions for segmentation while maintaining overall image fidelity. By incorporating attention mechanisms, feature fusion from multiple scales, and joint optimization with segmentation objectives, the reconstruction branch not only reconstructs the input image but also enhances the segmentation process by providing additional guidance and regularization. This integration improves the model's ability to capture meaningful representations, mitigate overfitting, and achieve better segmentation performance by ensuring that reconstructed images emphasize features crucial for accurate segmentation. Thus, this adapted reconstruction branch within the segmentation model framework offers improved interpretability, regularization, and segmentation accuracy compared to standalone reconstruction models.

\subsection{Upsampling Attention Gate Function:}
The upsampling attention gate described in Algorithm \ref{alg:upsample_attention_gate} operates by selectively attending to different regions of the input features while enhancing spatial details during upsampling tasks within neural network architectures. Initially, the input and context features undergo separate transformations via convolutional layers, followed by batch normalization to stabilize training. These transformed features are then combined and processed through activation functions to compute an attention map, indicating the relative importance of each spatial location. Crucially, the attention map is upsampled to match the spatial resolution of the input features, ensuring that attention weights are applied appropriately across all spatial dimensions. By multiplying the input features with the upsampled attention map, the upsampling attention gate selectively amplifies or suppresses features, effectively highlighting relevant regions and preserving spatial details during the upsampling process. This mechanism facilitates the reconstruction of high-resolution feature maps with enhanced fidelity, making it particularly advantageous for tasks such as image segmentation, where maintaining fine details is critical for performance. Attention mechanisms allow for selective feature emphasis, enhancing the model's ability to focus on relevant image regions while suppressing noise and irrelevant information. Additionally, upsampling residual blocks facilitates better gradient flow and feature propagation through skip connections, enabling the model to capture richer hierarchical representations and learn more discriminative features. Furthermore, AHNet dynamically adjusts the number of filters within residual blocks to match the dimensionality of the feature maps, enhancing adaptability and improving feature representation learning. Through these enhancements, AHNet performs superior medical image segmentation tasks by capturing long-range dependencies and contextual information more effectively, resulting in improved segmentation accuracy, robustness, and generalization across diverse datasets compared to traditional UNet architectures.

The key difference between the upsampling attention gate and a normal attention gate lies in their applicability and effectiveness in tasks involving upsampling operations within neural network architectures. While a normal attention gate focuses on selectively attending to different regions of the input features, the upsampling attention gate is specifically tailored to enhance upsampling tasks by effectively capturing and propagating spatial information at higher resolutions. By incorporating upsampling operations to match the size of the input features, the upsampling attention gate ensures that attention weights are applied at the exact spatial resolution of the input features, which is crucial for reconstructing high-resolution feature maps. This approach allows the upsampling attention gate to preserve spatial details better and improve the performance of tasks such as image segmentation, where maintaining fine details is essential. Thus, the upsampling attention gate offers a more targeted and effective solution for upsampling tasks compared to a normal attention gate.
%----------------------------------------
\section{Experiments}
\label{Experiments}
%-----------------------------------
\subsection{Dataset}
\label{Dataset}

The data is stored in the NIfTI (Neuroimaging Informatics Technology Initiative) format and comprises volumes of interest (VOIs) cropped around individual annotated lesions with a largest axial diameter of $5mm$ or greater. Each VOI is centered on a randomly selected lesion foreground voxel and is sized $256\times256\times128$ voxels in the original scan spacing. Padding was applied before cropping, using the minimum intensity value of the volume minus one.

A new data annotation approach, ULS23\_DeepLesion3D, was employed. In this method, trained biomedical students utilized measurement information from DeepLesion for 3D segmentation in the axial plane. Each lesion was segmented three times, and most masks were adopted as the final label. Lesions were chosen based on their performance compared to reference measurements from DeepLesion, with lesions exhibiting the poorest performance included. The selection aimed to represent the entire thorax-abdomen area, encompassing abdominal, bone, kidney, liver, lung, mediastinal, and assorted lesions.

Additionally, two specific datasets for bone and pancreas were derived from studies at Radboudumc Hospital in Nijmegen, The Netherlands. These datasets contain VOIs selected based on radiological reports mentioning bone or pancreas disease. An experienced radiologist identified and segmented the lesions in 3D. The bone dataset encompasses both sclerotic and lytic bone lesions.

\subsection{Data Preprocessing}
\label{Data_Preprocessing}
First, we extract image slices from volumetric medical image data encapsulated within NIfTI files. This task involves a series of intricate steps: normalization of each slice, conversion to grayscale PNG format, and resizing to a standardized dimension of $256\times 256$ pixels using the Lanczos interpolation technique. 

Subsequently, attention shifts to harmonizing image and label pairs, a pivotal step in the preprocessing pipeline. Then seamlessly amalgamate corresponding image and label data by meticulously matching filenames from distinct directories based on a shared identifier. The resulting fusion is encapsulated within HDF5 files, offering a structured and efficient means of storage conducive to the demands of subsequent deep learning endeavors. 
% Optionally, the code affords the flexibility to resize images before their consolidation, enhancing adaptability to diverse dataset characteristics.

In tandem, we orchestrate the systematic processing of images in batches, a strategy essential for handling large-scale datasets with finesse. Leveraging dictionaries as organizational scaffolds, the code systematically categorizes images by a unique identifier extracted from filenames. This methodical approach ensures the seamless association of each image with its pertinent metadata, thereby facilitating subsequent analysis and interpretation.

Lastly, we use patient identities to organize the data. The new HDF5 files are customized to each patient's unique ID by navigating the combined dataset repeatedly. Image and label data relating to specific patients reside in these custom archives, providing a well-organized and cohesive framework that is beneficial for further research and model training efforts.

\subsection{Evaluation Metrics}
\label{Evaluation_Metrics}
The evaluation of segmentation algorithms in this study involved metrics such as the dice similarity coefficient (DSC), intersection over union (IoU), mean Hausdorff distance (MHD), relative absolute volume difference (RAVD), average surface distance (ASD) and area under receiver operating characteristic (ROC) curve (AUC).

% The CCE quantifies the difference between the predicted output and the ground truth. In image segmentation tasks, where the model aims to classify each pixel into different classes that’s why we used CCE. This CCE is appropriate when the classes are mutually exclusive and each pixel should belong to only one class. Although we use CCE curves to measure whether the model is overfit, underfit or good fit. The following formula gives the CCE function:
% \begin{equation*}
%     CCE=-\frac{1}{N}{\sum_{i=1}^N log p_{model}[y_i\in C_{y_i}]}
% \end{equation*}
% where ${N}$ represents the total number of pixels or data points, ${y_i}$ denotes the ground truth label for pixel ${i}$, ${C_{y_i}}$ signifies the set of possible classes for pixel ${i}$ and ${p_{model}[y_i\in C_{y_i}]}$ represents the predicted probability assigned by the model to the ground truth class of pixel ${i}$. Model accuracy is a pivotal metric in image segmentation, assessing the ratio of correctly classified pixels to the total number of pixels in the medical images. This metric provides a fundamental evaluation of how accurately the algorithm identifies the region. 
% \begin{equation*}
%     Accuracy=\frac{Number \: of \: correct \: predictions}{Total \: number \: of \: predictions}
% \end{equation*}
% Also, to evaluate the performance of the prostate anatomy and tumor segmentation algorithm, the DSC is analyzed. 

We utilized the DSC as an area-based metric to quantify the degree of spatial overlap between the predicted mask and the ground truth. This metric offers a precise measure of segmentation accuracy. Also, we present the IoU (Jaccard Index) for the segmentation tasks, to ensure that the annotations that are delivered have a mean $IoU$ ${\geq X}$ (where ${X = 0.5}$) concerning the “good” segmentation.
\begin{equation*}
    DSC=\frac{2 \times TP}{2 \times TP + FP + FN}
\end{equation*}
\begin{align*}
    IoU &= \frac{Intersection \: Area}{Union \: Area}\\
        &= \frac{TP}{TP+FP+FN}
\end{align*}
where, TP (True Positives) represents the number of pixels correctly classified as belonging to the region of interest, FP (False Positives) represents the number of pixels in the predicted mask that do not correspond to the region of interest, FN (False Negatives) represents the number of pixels in the ground truth mask that were missed by the segmentation algorithm. Besides, AUC is a crucial segmentation metric that summarizes the model's overall performance across all possible threshold settings. A perfect classifier would have an AUC of 1, indicating perfect discrimination between positive and negative samples. Specifically, AUC is calculated from the ROC curve, a graphical plot that illustrates the true positive rate (sensitivity) against the false positive rate (1-specificity) at various threshold settings. It provides insights into the segmentation model's performance in distinguishing positive and negative regions.

% Also, to calculate the proportion of true prostate pixels correctly identified by the segmentation algorithm, emphasizing the ability to detect actual positive instances we implement sensitivity, also known as Recall which is defined as,
% \begin{equation*}
%     Sensitivity=\frac{TP}{TP + FN}
% \end{equation*}
% On the other hand, specificity quantifies the proportion of non-prostate pixels correctly identified as non-prostate regions by the algorithm. It highlights the algorithm's effectiveness in recognizing negative instances.
% \begin{equation*}
%     Specificity=\frac{TN}{TN + FP}
% \end{equation*}
% To measure the ratio of pixels within the actual prostate region that are erroneously classified as non-prostate. This metric FNR provides insights into potential under-segmentation issues. But FPR assesses the ratio of pixels outside the true prostate region that are mistakenly classified as prostate which highlights potential over-segmentation challenges.

% \begin{equation*}
%     FNR=\frac{FN}{TP + FN}
% \end{equation*}
% \begin{equation*}
%     FPR=\frac{FP}{TN + FP}
% \end{equation*}
% In the results we will also present the IoU (Jaccard Index) for their prostate segmentation tasks, to ensure that the annotations that are delivered have a mean $IoU$ ${\geq X}$ (where ${X = 0.5}$) with respect to the “good” segmentation.
% \begin{align*}
%     IoU &= \frac{Intersection \: Area}{Union \: Area}\\
%         &= \frac{TP}{TP+FP+FN}
% \end{align*}
% where ${TP}$ represents true positive, ${TN}$ represents true negative, ${FP}$ represents false positive and ${FN}$ represents false negative. 

MHD provides insight into the average maximum distance between points in the predicted mask and ground truth. This metric offers a comprehensive evaluation of boundary discrepancies which is followed by, 
\begin{equation*}
    MHD=\frac{1}{N} \sum_{i=1}^N \max_j \min_k d(P_i, T_j)
\end{equation*}
where \(N\) represents the total number of data points, \(P_i\) is a point in the predicted mask, \(T_j\) is a point in the ground truth and \(d(P_i, T_j)\) calculates the distance between \(P_i\) and \(T_j\). The metric RAVD is particularly relevant for understanding volume-related disparities.
\begin{equation*}
    RAVD=\frac{|V_P - V_T|}{V_T}
\end{equation*}
Here \(V_P\) represents the volume of the predicted mask and \(V_T\) represents the volume of the ground truth. Also, ASD calculates the average distance between the predicted mask and ground truth surfaces, focusing on boundary accuracy.

% \begin{equation*}
%     ASSD=\frac{1}{N} \sum_{i=1}^N \frac{1}{|P_i|} \sum_{p \in P_i} \min_{q \in T} \left[d(p, q) + d(q, p)\right]
% \end{equation*}

\begin{equation*}
    ASD=\frac{1}{N} \sum_{i=1}^N \frac{1}{|P_i|} \sum_{p \in P_i} \min_{q \in T} d(p, q)
\end{equation*}
where \(N\) represents the total number of data points, \(P_i\) is the predicted mask for data point \(i\), \(|P_i|\) denotes the number of points on \(P_i\), \(p\) is a point on \(P_i\), \(T\) represents the ground truth and \(d(p, q)\) calculates the distance between point \(p\) on the predicted mask and point \(q\) on the ground truth.

\section{Computational Time}
\label{Computation_Time}
This section analyzes the computational time associated with implementing the proposed Mamba-AHNet with image reconstruction. The experiments were conducted on a machine equipped with the following specifications:
\begin{itemize}
    \item \textbf{Processors: }AMD Ryzen Threadripper 1920X 12-Core processor
    \item \textbf{Clock Speed: }3.5 GHz
    \item \textbf{RAM: }32 GB
    \item \textbf{Environment: }Python 3.9 with Tensorflow framework
\end{itemize}

\begin{table}[]
\setlength{\tabcolsep}{10.4pt} % Column gap
\renewcommand{\arraystretch}{1.5} % Row gap
\begin{tabular}{llll}
\hline
Dataset &
  Samples &
  Model &
  \begin{tabular}[l]{@{}l@{}}Training \\ time (s) \\ (per epoch)\end{tabular} \\ \hline
 &
   &
  HUNet &
  1997.1423 \\ 
 &
   &
  \cellcolor[HTML]{DEF0F9}AHNet &
  \cellcolor[HTML]{DEF0F9}3678.7891 \\ 
 &
   &
  \begin{tabular}[c]{@{}l@{}}AHNet \\ with image \\ reconstruction\end{tabular} &
  3712.1234 \\ 
 &
   &
  \cellcolor[HTML]{DEF0F9}Mamba-AHNet &
  \cellcolor[HTML]{DEF0F9}2162.8765 \\ 
\multirow{-5}{*}{Deeplesion} &
  \multirow{-5}{*}{\begin{tabular}[c]{@{}l@{}}24580 (Train)\\ 6145 (Test)\\ 3414 (Val)\end{tabular}} &
  \begin{tabular}[c]{@{}l@{}}Mamba-AHNet \\ with image \\ reconstruction\end{tabular} &
  2302.3456 \\ \hline
 &
   &
  \cellcolor[HTML]{DEF0F9}HUNet &
  \cellcolor[HTML]{DEF0F9}5295.8901 \\ 
 &
   &
  AHNet &
  10934.5678 \\ 
 &
   &
  \cellcolor[HTML]{DEF0F9}\begin{tabular}[c]{@{}l@{}}AHNet \\ with image \\ reconstruction\end{tabular} &
  \cellcolor[HTML]{DEF0F9}12987.2345 \\  
 &
   &
  Mamba-AHNet &
  5483.9012 \\  
\multirow{-5}{*}{\begin{tabular}[c]{@{}l@{}}ULS \\ Bone\end{tabular}} &
  \multirow{-5}{*}{\begin{tabular}[c]{@{}l@{}}62339 (Train)\\ 15585 (Test)\\ 8659 (Val)\end{tabular}} &
  \cellcolor[HTML]{DEF0F9}\begin{tabular}[c]{@{}l@{}}Mamba-AHNet \\ with image \\ reconstruction\end{tabular} &
  \cellcolor[HTML]{DEF0F9}5892.6789 \\ \hline
 &
   &
  HUNet &
  746.9876 \\ 
 &
   &
  \cellcolor[HTML]{DEF0F9}AHNet &
  \cellcolor[HTML]{DEF0F9}1728.4567 \\ 
 &
   &
  \begin{tabular}[c]{@{}l@{}}AHNet \\ with image \\ reconstruction\end{tabular} &
  1918.7890 \\ 
 &
   &
  \cellcolor[HTML]{DEF0F9}Mamba-AHNet &
  \cellcolor[HTML]{DEF0F9}859.1098 \\ 
\multirow{-5}{*}{\begin{tabular}[c]{@{}l@{}}ULS \\ Pancreas\end{tabular}} &
  \multirow{-5}{*}{\begin{tabular}[c]{@{}l@{}}9760 (Train)\\ 2440 (Test)\\ 1356 (Val)\end{tabular}} &
  \begin{tabular}[c]{@{}l@{}}Mamba-AHNet \\ with image \\ reconstruction\end{tabular} &
  1025.8764 \\ \hline
\end{tabular}
\caption{Computational time and CPU usage for both training and testing phases for each dataset using HUNet, AHNet, AHNet with image reconstruction, Mamba-AHNet, and Mamba-AHNet with image reconstruction (times are measured in seconds)}
    \label{tab:computational_time}
\end{table}

Across all datasets, the training time for Mamba-AHNet with image reconstruction was recorded and compared against other models in Table \ref{tab:computational_time}. Notably, training times varied significantly depending on the neural network's architecture and image reconstruction techniques' presence. For instance, in the Deeplesion dataset, HUNet exhibited the shortest training time per epoch at 1997.1423 seconds, while AHNet with image reconstruction required the longest time at 3712.1234 seconds. Similarly, in the ULS Bone dataset, HUNet's training time per epoch was comparatively shorter than AHNet with image reconstruction.

A notable trade-off emerges when considering the incorporation of image reconstruction techniques. While these techniques may enhance the model's performance, they often incur a computational cost. This is evident in the case of AHNet with image reconstruction and Mamba-AHNet with image reconstruction, both of which demonstrated longer training times than their counterparts without image reconstruction.

Despite the substantial computational requirements associated with incorporating image reconstruction techniques, Mamba-AHNet with image reconstruction showed competitive performance in training time and model effectiveness.

It is observed that Mamba-AHNet with image reconstruction incurred longer training times than its counterpart without image reconstruction. However, this increase in computational overhead was offset by the superior performance achieved by the model.

\section{Numerical Analysis}
\label{Numerical_Analysis}
\subsection{Results}
\label{Results}
%%%%%%%%%%%%%%%%%%%%%%%%%%%%%%%%%
\begin{figure}[htb!]
\centering
\begin{tabular}{ccc}
 Test Image & Test Label & Predicted Label \\
    \includegraphics[width=0.29\linewidth]{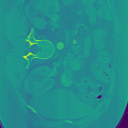} &  
    \begin{tikzpicture}
        % Include original image
        \node[anchor=south west,inner sep=0] (image) at (0,0) {\includegraphics[width=0.29\linewidth]{Figures/d10.png}};
        % Include original mask with 40% transparency
        \node[anchor=south west,inner sep=0,opacity=1] (mask) at (0,0) {\includegraphics[width=0.29\linewidth]{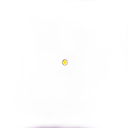}};
    \end{tikzpicture} & 
    \begin{tikzpicture}
        % Include original image
        \node[anchor=south west,inner sep=0] (image) at (0,0) {\includegraphics[width=0.29\linewidth]{Figures/d10.png}};
        % Include original mask with 40% transparency
        \node[anchor=south west,inner sep=0,opacity=1] (mask) at (0,0) {\includegraphics[width=0.29\linewidth]{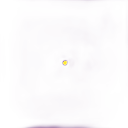}};
    \end{tikzpicture}  \\
    \includegraphics[width=0.29\linewidth]{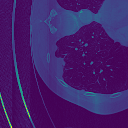} &  
    \begin{tikzpicture}
        % Include original image
        \node[anchor=south west,inner sep=0] (image) at (0,0) {\includegraphics[width=0.29\linewidth]{Figures/d20.png}};
        % Include original mask with 40% transparency
        \node[anchor=south west,inner sep=0,opacity=1] (mask) at (0,0) {\includegraphics[width=0.29\linewidth]{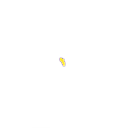}};
    \end{tikzpicture} & 
    \begin{tikzpicture}
        % Include original image
        \node[anchor=south west,inner sep=0] (image) at (0,0) {\includegraphics[width=0.29\linewidth]{Figures/d20.png}};
        % Include original mask with 40% transparency
        \node[anchor=south west,inner sep=0,opacity=1] (mask) at (0,0) {\includegraphics[width=0.29\linewidth]{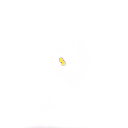}};
    \end{tikzpicture}  \\
\end{tabular}
\caption{Output segmentation results for deeplesion dataset}
\label{fig:segmented_results_deeplesion}
\end{figure}
%%%%%%%%%%%%%%%%%%%%%%%%%%%%%%%%%

%%%%%%%%%%%%%%%%%%%%%%%%%%%%%%%%%
\begin{figure}[htb!]
\centering
\begin{tabular}{ccc}
 Test Image & Test Label & Predicted Label \\
    \includegraphics[width=0.29\linewidth]{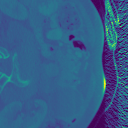} &  
    \begin{tikzpicture}
        % Include original image
        \node[anchor=south west,inner sep=0] (image) at (0,0) {\includegraphics[width=0.29\linewidth]{Figures/p10.png}};
        % Include original mask with 40% transparency
        \node[anchor=south west,inner sep=0,opacity=1] (mask) at (0,0) {\includegraphics[width=0.29\linewidth]{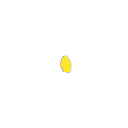}};
    \end{tikzpicture} & 
    \begin{tikzpicture}
        % Include original image
        \node[anchor=south west,inner sep=0] (image) at (0,0) {\includegraphics[width=0.29\linewidth]{Figures/p10.png}};
        % Include original mask with 40% transparency
        \node[anchor=south west,inner sep=0,opacity=1] (mask) at (0,0) {\includegraphics[width=0.29\linewidth]{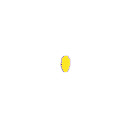}};
    \end{tikzpicture}  \\
    \includegraphics[width=0.29\linewidth]{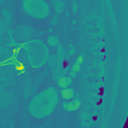} &  
    \begin{tikzpicture}
        % Include original image
        \node[anchor=south west,inner sep=0] (image) at (0,0) {\includegraphics[width=0.29\linewidth]{Figures/p20.png}};
        % Include original mask with 40% transparency
        \node[anchor=south west,inner sep=0,opacity=0] (mask) at (0,0) {\includegraphics[width=0.29\linewidth]{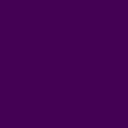}};
    \end{tikzpicture} & 
    \begin{tikzpicture}
        % Include original image
        \node[anchor=south west,inner sep=0] (image) at (0,0) {\includegraphics[width=0.29\linewidth]{Figures/p20.png}};
        % Include original mask with 40% transparency
        \node[anchor=south west,inner sep=0,opacity=0] (mask) at (0,0) {\includegraphics[width=0.29\linewidth]{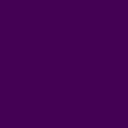}};
    \end{tikzpicture}  \\
\end{tabular}
\caption{Output segmentation results for pancreas dataset}
\label{fig:segmented_results_pancreas}
\end{figure}
%%%%%%%%%%%%%%%%%%%%%%%%%%%%%%%%%

%%%%%%%%%%%%%%%%%%%%%%%%%%%%%%%%%
\begin{figure}[htb!]
\centering
\begin{tabular}{ccc}
 Test Image & Test Label & Predicted Label \\
    \includegraphics[width=0.29\linewidth]{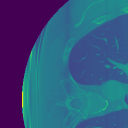} &  
    \begin{tikzpicture}
        % Include original image
        \node[anchor=south west,inner sep=0] (image) at (0,0) {\includegraphics[width=0.29\linewidth]{Figures/b10.png}};
        % Include original mask with 40% transparency
        \node[anchor=south west,inner sep=0,opacity=1] (mask) at (0,0) {\includegraphics[width=0.29\linewidth]{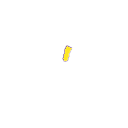}};
    \end{tikzpicture} & 
    \begin{tikzpicture}
        % Include original image
        \node[anchor=south west,inner sep=0] (image) at (0,0) {\includegraphics[width=0.29\linewidth]{Figures/b10.png}};
        % Include original mask with 40% transparency
        \node[anchor=south west,inner sep=0,opacity=1] (mask) at (0,0) {\includegraphics[width=0.29\linewidth]{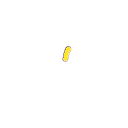}};
    \end{tikzpicture}  \\
    \includegraphics[width=0.29\linewidth]{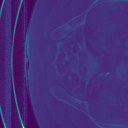} &  
    \begin{tikzpicture}
        % Include original image
        \node[anchor=south west,inner sep=0] (image) at (0,0) {\includegraphics[width=0.29\linewidth]{Figures/b20.png}};
        % Include original mask with 40% transparency
        \node[anchor=south west,inner sep=0,opacity=1] (mask) at (0,0) {\includegraphics[width=0.29\linewidth]{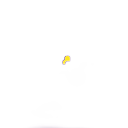}};
    \end{tikzpicture} & 
    \begin{tikzpicture}
        % Include original image
        \node[anchor=south west,inner sep=0] (image) at (0,0) {\includegraphics[width=0.29\linewidth]{Figures/b20.png}};
        % Include original mask with 40% transparency
        \node[anchor=south west,inner sep=0,opacity=1] (mask) at (0,0) {\includegraphics[width=0.29\linewidth]{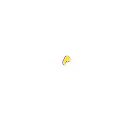}};
    \end{tikzpicture}  \\
\end{tabular}
\caption{Output segmentation results for bone dataset}
\label{fig:segmented_results_bone}
\end{figure}
%%%%%%%%%%%%%%%%%%%%%%%%%%%%%%%%%

%%%%%%%%%%%%%%%%table1%%%%%%%%%%%%%%%%%%%%%%%%%
\begin{table*}[]
\setlength{\tabcolsep}{16.5pt} % Column gap
\renewcommand{\arraystretch}{1.5} % Row gap
\begin{tabular}{lcccccc}
\hline
\textbf{Study} &
  \textbf{DSC$\uparrow$} &
  \textbf{RAVD$\downarrow$} &
  \textbf{ASD$\downarrow$} &
  \textbf{MHD$\downarrow$} &
  \textbf{AUC$\uparrow$} &
  \textbf{IoU$\uparrow$} \\ \hline
 \textbf{MULAN} \cite{yan2019mulan} & 0.7534 & 0.7120 & 0.1269 & 0.5839 & 0.9601 & 0.6785 \\
\rowcolor[HTML]{DEF0F9} 
 \textbf{ULD} \cite{lyu2021segmentation} & 0.7987 & 0.5784 & 0.1070 & 0.5169 & 0.9641 & 0.6943 \\
     \textbf{ULS4US} \cite{wu2023uls4us} & 0.8912 & 0.4031 & 0.0641 & 0.4206 & 0.9781 & 0.7310 \\ 
\rowcolor[HTML]{DEF0F9} 
 \textbf{C2FVL} \cite{shan2023coarse} & 0.8907 & 0.3893 & 0.0767 & 0.4594 & 0.9774 & 0.7116 \\
     \textbf{DCAU-Net} \cite{yuan2022dcau} & 0.8164 & 0.4135 & 0.0839 & 0.4798 & 0.9685 & 0.6995 \\ 
\rowcolor[HTML]{DEF0F9} 
 \textbf{HUNet}  & 0.8789 & 0.3766 & 0.0669 & 0.4122 & 0.9750 & 0.7016 \\
     \textbf{AHNet}  & 0.9676 & 0.2776 & 0.0467 & 0.3192 & 0.9879 & 0.7944 \\
\rowcolor[HTML]{DEF0F9} 
 \textbf{AHNet with image reconstruction}  & 0.9821 & 0.2574 & 0.0319 & 0.2167 & 0.9891 & 0.8144 \\
     \textbf{Mamba-AHNet}  & 0.9874 & 0.2688 & 0.0265 & 0.2128 & 0.9952 & 0.8092 \\
\rowcolor[HTML]{DEF0F9} 
 \textbf{Mamba-AHNet with image reconstruction}  & \textbf{0.9896} & \textbf{0.2452} & \textbf{0.0140} & \textbf{0.1999} & \textbf{0.9963} & \textbf{0.8265} \\ \hline
\end{tabular}
\caption{Performance of different models on Deeplesion dataset. $\uparrow$ denotes higher is better and $\downarrow$ denotes lower is better}
    \label{tab:table1}
\end{table*}
%%%%%%%%%%%%%%%%%%table1%%%%%%%%%%%%%%%%%%%%%%
%%%%%%%%%%%%%%%%%%table2%%%%%%%%%%%%%%%%%%%%%%
\begin{table*}[]
\setlength{\tabcolsep}{16.5pt} % Column gap
\renewcommand{\arraystretch}{1.5} % Row gap
\begin{tabular}{lcccccc}
\hline
\textbf{Study} &
  \textbf{DSC$\uparrow$} &
  \textbf{RAVD$\downarrow$} &
  \textbf{ASD$\downarrow$} &
  \textbf{MHD$\downarrow$} &
  \textbf{AUC$\uparrow$} &
  \textbf{IoU$\uparrow$} \\ \hline
 \textbf{MULAN} \cite{yan2019mulan} & 0.7398 & 0.7354 & 0.1405 & 0.5911 & 0.9562 & 0.6605 \\ 
\rowcolor[HTML]{DEF0F9} 
 \textbf{ULD} \cite{lyu2021segmentation} & 0.8059 & 0.5521 & 0.1008 & 0.5362 & 0.9613 & 0.6832 \\ 
 \textbf{ULS4US} \cite{wu2023uls4us} & 0.8831 & 0.3845 & 0.0738 & 0.4098 & 0.9752 & 0.7423 \\
\rowcolor[HTML]{DEF0F9} 
 \textbf{C2FVL} \cite{shan2023coarse} & 0.8655 & 0.4036 & 0.0925 & 0.4512 & 0.9721 & 0.7306 \\
    \textbf{DCAU-Net} \cite{yuan2022dcau} & 0.8276 & 0.4059 & 0.0813 & 0.4845 & 0.9663 & 0.7063 \\ 
\rowcolor[HTML]{DEF0F9} 
 \textbf{HUNet}  & 0.8637 & 0.3675 & 0.0898 & 0.4603 & 0.9687 & 0.7128 \\
    \textbf{AHNet}  & 0.9635 & 0.2664 & 0.0489 & 0.2976 & 0.9855 & 0.8063 \\ 
\rowcolor[HTML]{DEF0F9} 
 \textbf{AHNet with image reconstruction}  & 0.9776 & 0.2558 & 0.0337 & 0.2335 & 0.9873 & 0.8225 \\
    \textbf{Mamba-AHNet}  & 0.9802 & 0.2498 & 0.0286 & 0.2034 & 0.9935 & 0.8261 \\ 
\rowcolor[HTML]{DEF0F9} 
\textbf{Mamba-AHNet with image reconstruction}  & \textbf{0.9820} & \textbf{0.2365} & \textbf{0.0178} & \textbf{0.1876} & \textbf{0.9947} & \textbf{0.8339} \\ \hline
\end{tabular}
\caption{Performance of different models on ULS Bone dataset. $\uparrow$ denotes higher is better and $\downarrow$ denotes lower is better}
    \label{tab:table2}
\end{table*}
%%%%%%%%%%%%%%%%%%table2%%%%%%%%%%%%%%%%%%%%%%
%%%%%%%%%%%%%%%%%%table3%%%%%%%%%%%%%%%%%%%%%%
\begin{table*}[]
\setlength{\tabcolsep}{16.5pt} % Column gap
\renewcommand{\arraystretch}{1.5} % Row gap
\begin{tabular}{lcccccc}
\hline
\textbf{Study} &
  \textbf{DSC$\uparrow$} &
  \textbf{RAVD$\downarrow$} &
  \textbf{ASD$\downarrow$} &
  \textbf{MHD$\downarrow$} &
  \textbf{AUC$\uparrow$} &
  \textbf{IoU$\uparrow$} \\ \hline
 \textbf{MULAN} \cite{yan2019mulan} & 0.7456 & 0.7296 & 0.1348 & 0.5672 & 0.9573 & 0.6659 \\ 
\rowcolor[HTML]{DEF0F9} 
 \textbf{ULD} \cite{lyu2021segmentation} & 0.8014 & 0.5623 & 0.0945 & 0.5267 & 0.9627 & 0.6889 \\
    \textbf{ULS4US} \cite{wu2023uls4us} & 0.8775 & 0.3887 & 0.0695 & 0.4156 & 0.9765 & 0.7378 \\ 
\rowcolor[HTML]{DEF0F9} 
 \textbf{C2FVL} \cite{shan2023coarse} & 0.8702 & 0.3998 & 0.0862 & 0.4392 & 0.9732 & 0.7251 \\
    \textbf{DCAU-Net} \cite{yuan2022dcau} & 0.8221 & 0.4097 & 0.0776 & 0.4813 & 0.9671 & 0.7024 \\
\rowcolor[HTML]{DEF0F9} 
 \textbf{HUNet}  & 0.8585 & 0.4133 & 0.0716 & 0.4215 & 0.9716 & 0.7203 \\
    \textbf{AHNet}  & 0.9712 & 0.2701 & 0.0512 & 0.3065 & 0.9863 & 0.8017 \\ 
\rowcolor[HTML]{DEF0F9} 
 \textbf{AHNet with image reconstruction}  & 0.9868 & 0.2613 & 0.0356 & 0.2256 & 0.9882 & 0.8178 \\
    \textbf{Mamba-AHNet}  & 0.9852 & 0.2531 & 0.0307 & 0.2073 & 0.9941 & 0.8229 \\
\rowcolor[HTML]{DEF0F9} 
 \textbf{Mamba-AHNet with image reconstruction}  & \textbf{0.9891} & \textbf{0.2397} & \textbf{0.0193} & \textbf{0.1912} & \textbf{0.9952} & \textbf{0.8306} \\ \hline
\end{tabular}
\caption{Performance of different models on ULS Pancreas dataset. $\uparrow$ denotes higher is better and $\downarrow$ denotes lower is better}
    \label{tab:table3}
\end{table*}
%%%%%%%%%%%%%%%%%%table3%%%%%%%%%%%%%%%%%%%%%%

\begin{figure*}[t]
     \centering
     \includegraphics[width=\textwidth]{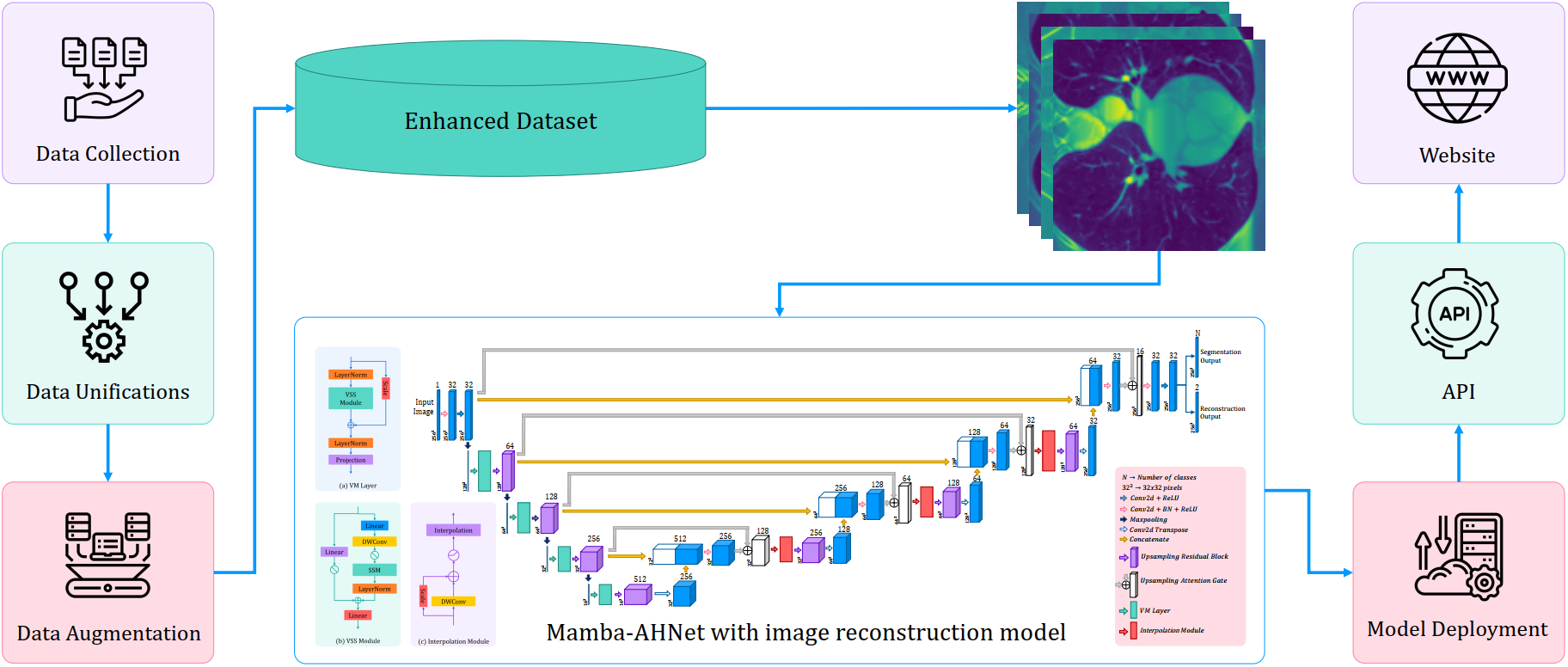}
     \caption{Workflow diagram illustrating the process of data collection, unifications, augmentation, and model deployment in the development of Mamba-AHNet with image reconstruction model for enhanced dataset analysis and integration into a website through an API}
     \label{fig:website}
\end{figure*}

In Table \ref{tab:table1}, we comprehensively compare various models' performance on the Deeplesion3D dataset. Upon scrutiny, it becomes apparent that Mamba-AHNet with image reconstruction outperforms other models, including HUNet, AHNet, AHNet with image reconstruction, MAMBA-AHNet, and baseline models  \cite{shan2023coarse, yuan2022dcau, lyu2021segmentation, wu2023uls4us, yan2019mulan}. Mamba-AHNet with image reconstruction achieves a DSC score of 0.9896, indicating a superior overlap between predicted and ground truth segmentations compared to the baseline models. Additionally, it demonstrates lower values for RAVD, ASD, and MHD, indicating improved segmentation accuracy and spatial alignment. Furthermore, Mamba-AHNet with image reconstruction attains an impressive AUC of 0.9963 and an IoU of 0.8265 on the Deeplesion3D dataset, underscoring its exceptional capability in accurately delineating lesions within medical images.

Moving to Table \ref{tab:table2}, which presents findings on the ULS Bone dataset, Mamba-AHNet with image reconstruction continues to demonstrate superiority over the baseline models. While the baseline models show DSC scores as low as 0.7398, Mamba-AHNet with image reconstruction achieves a remarkable DSC of 0.9820, indicating a substantial enhancement in bone structure segmentation accuracy. Furthermore, Mamba-AHNet with image reconstruction exhibits lower values for RAVD, ASD, and MHD than the baseline models, emphasizing its effectiveness in precisely delineating bone structures in medical images. Additionally, it achieves an AUC of 0.9947 and an IoU of 0.8339, further highlighting its proficiency in accurately segmenting bone structures.

Likewise, Table \ref{tab:table3} provides insights into the performance comparison on the ULS Pancreas dataset, where Mamba-AHNet with image reconstruction showcases remarkable performance. While the baseline models achieve minimal DSC scores, Mamba-AHNet with image reconstruction achieves a noteworthy DSC of 0.9891, significantly improving pancreas segmentation accuracy. Moreover, Mamba-AHNet with image reconstruction demonstrates lower values for RAVD, ASD, MHD, and IoU than the baseline models, confirming its efficacy in precisely delineating pancreas structures in medical images. Additionally, it achieves an AUC of 0.9952 and an IoU of 0.8306, further emphasizing its effectiveness in accurately delineating pancreas structures within medical images. High AUC and IoU values underscore the precise segmentation and localization of pancreas regions achieved by Mamba-AHNet with image reconstruction.

To complement the quantitative analysis, visual representations of the segmentation results for each dataset were provided in Figure \ref{fig:segmented_results_deeplesion}, \ref{fig:segmented_results_pancreas} and \ref{fig:segmented_results_bone}. These visualizations offer a qualitative assessment of the segmentation performance of Mamba-AHNet with image reconstruction compared to the baseline models, further supporting the conclusions drawn from the quantitative analysis.

\subsection{Discussion}
\label{Discussion}
The results obtained from the comparative analysis demonstrate the superior performance of Mamba-AHNet with image reconstruction in medical image segmentation tasks across multiple datasets. Across all three datasets, Mamba-AHNet with image reconstruction consistently outperformed the baseline models in terms of DSC, RAVD, ASD, MHD, AUC, and IoU. These findings underscore the efficacy and robustness of Mamba-AHNet with image reconstruction in accurately delineating anatomical structures in medical images, thus addressing the primary objective of the research.

Interpreting the results reveals several key insights. Firstly, the significantly higher DSC scores, AUC, and IoU achieved by Mamba-AHNet with image reconstruction indicate a better overlap between the predicted and ground truth segmentations, emphasizing its superior segmentation accuracy. Additionally, the lower values for RAVD, ASD and MHD further confirm the enhanced precision and spatial alignment of Mamba-AHNet with image reconstruction segmentations compared to the baseline models. These interpretations highlight the effectiveness of the proposed model in addressing the challenges associated with medical image segmentation, such as accurate localization and delineation of anatomical structures.

The implications of these results are profound for both clinical practice and research. The superior performance of Mamba-AHNet with image reconstruction holds great promise for various medical applications, including disease diagnosis, treatment planning, and monitoring. Accurate and precise segmentation of medical images is crucial for clinical decision-making, and the efficacy of Mamba-AHNet with image reconstruction signifies its potential to improve patient outcomes and streamline healthcare workflows. Moreover, the robustness of Mamba-AHNet with image reconstruction across different datasets underscores its versatility and generalizability, further enhancing its utility in diverse clinical scenarios.

Additionally, to facilitate the utilization of Mamba-AHNet with image reconstruction in clinical settings, a user-friendly website has been developed. This website is connected via API, providing seamless integration with existing medical imaging systems. Radiologists can easily access the functionality of the model through this interface, enabling efficient segmentation of medical images. Furthermore, due to the lightweight nature of the model, the response from the API is fast, ensuring quick turnaround times for image analysis. Also, the workflow of the developed system in Figure \ref{fig:website}, highlights the integration of Mamba-AHNet with image reconstruction into the medical imaging pipeline and its interaction with the user-friendly website via API.

\section{Conclusion}
\label{Conclusion}
In conclusion, integrating the SSM with the AHNet within the MAMBA framework, as proposed in our methodology named Mamba-AHNet, demonstrates remarkable efficacy in addressing the challenges of semantic segmentation in medical imaging. Through a combination of SSM's feature extraction capabilities, AHNet's attention mechanisms and residual blocks, our approach enhances segmentation accuracy and robustness.
The comprehensive evaluation of various datasets, including Deeplesion3D, ULS Bone, and ULS Pancreas, consistently illustrates the superiority of Mamba-AHNet with image reconstruction over baseline models in terms of metrics such as DSC, RAVD, ASD, MHD, AUC, and IoU. These findings highlight the model's ability to accurately delineate anatomical structures in medical images, which is crucial for diagnosis, treatment planning, and monitoring in clinical practice. Moreover, the qualitative assessment through visual representations reinforces the quantitative results, providing further evidence of the superior segmentation performance of Mamba-AHNet with image reconstruction. These findings' implications extend to clinical practice and research, promising advancements in disease diagnosis, treatment planning, and patient care. The versatility and generalizability of Mamba-AHNet with image reconstruction across diverse datasets underscore its potential to streamline healthcare workflows and improve patient outcomes. The proposed methodology represents a significant advancement in medical image segmentation, offering a robust and effective solution to the challenges encountered in clinical settings.

\section*{Authors Contribution}
\label{Authors Contribution}
Conceptualization: K S Sanjid, M T Hossain, M S S Junayed. Data curation: M T Hossain, K S Sanjid. Formal Analysis: M S S Junayed, K S Sanjid, M T Hossain. Methodology: K S Sanjid, M T Hossain, M S S Junayed. Supervision: M M Uddin. Writing – original draft: M T Hossain, M S S Junayed, K S Sanjid, M M Uddin. Funding acquisition: M M Uddin. K S Sanjid, M S S Junayed, and M T Hossain all made equal contributions and are entitled to have their names listed first on their CVs. All authors played a role in the article's development and approved the submitted version.

% \section*{Acknowledgements}
% The research is funded by the North South University Conference and Travel Grant Committee (ID: CTRG-22-SEPS-06)

\bibliography{bibliography}
\bibliographystyle{IEEEtran}
\end{document}